\documentclass[journal]{IEEEtran}
\pdfoutput=1

\ifCLASSINFOpdf
\else
\fi

\usepackage{array}
\usepackage{times}
\usepackage{epsfig}
\usepackage{graphicx}
\usepackage{amsmath}
\usepackage{amssymb}
\usepackage{graphicx}
\usepackage{cite}
\usepackage{enumerate}
\usepackage{cases}
\usepackage{multirow}
\usepackage{verbatim}
\usepackage{amssymb}
\usepackage{CJK}
\usepackage{algorithm}
\usepackage{algorithmicx}
\usepackage{algpseudocode}
\usepackage{stfloats}
\usepackage{color}

\usepackage{bm}
\usepackage{booktabs}
\usepackage[colorlinks,linkcolor=red]{hyperref}
\usepackage{cleveref} 
\usepackage{hyperref}
\usepackage{graphicx}

\begin{document}


\title{A Near-Field Super-Resolution Network for Accelerating Antenna Characterization}
\author{Yuchen Gu, \IEEEmembership{Student Member, IEEE}, Hai-Han Sun,  \IEEEmembership{Member, IEEE}, and Daniel W. van der Weide, \IEEEmembership{Fellow, IEEE}

\thanks{The authors are with the Electrical and Computer Engineering Department, University of Wisconsin-Madison, WI 53706, USA \break(e-mail: danvdw@engr.wisc.edu).}

}


\maketitle

\begin{abstract}

We present a deep neural network-enabled method to accelerate near-field (NF) antenna measurement. Coupled with a large synthetic dataset of antenna field maps and novel magnitude and phase loss functions, we develop a Near-field Super-resolution Network (NFS-Net) to reconstruct significantly undersampled near-field data into high-resolution data. This approach considerably reduces the number of sampling points required for NF measurement and thus improves measurement efficiency. The high-resolution near-field data reconstructed by the network is further processed by a near-field-to-far-field (NF2FF) transformation to obtain far-field antenna radiation patterns. Our experiments demonstrate that the NFS-Net exhibits both accuracy and generalizability in restoring high-resolution near-field data from low-resolution input. The NF measurement workflow that combines the NFS-Net and the NF2FF algorithm enables accurate radiation pattern characterization with only 11\% of the Nyquist rate samples. Though the experiments in this study are conducted on a planar setup with a uniform grid, the proposed method can serve as a universal strategy to accelerate measurements under different setups and conditions.

\end{abstract}


\IEEEpeerreviewmaketitle

\section{Introduction}

\IEEEPARstart {N}{ear-field} (NF) antenna measurements are widely used to characterize the radiation patterns of antennas because of their accuracy and convenience of operation in a controlled environment \cite{Field-similarity}, \cite{Antennatesting_mmWave}. With modern communication devices increasingly operating at shorter (millimeter) wavelength, NF measurements have become even more favored to address significant path loss \cite{Planar-NF}. Compared to far-field (FF) antenna measurement techniques, NF measurement systems have the advantage of reducing the required testing distance and facility space, which is particularly beneficial for the characterization of electrically large antennas \cite{Field-similarity}. However, near-field measurements are usually time-consuming due to the requirement of sampling a large number of spatial coordinates, posing a burden on  the growing demand for efficient antenna characterization \cite{Planar-NF}, \cite{optimal-sampling}, \cite{pattern-rotation}, \cite{balanis}, \cite{planar_full_hemp}.  \par

Since data acquisition time is directly proportional to the number of field samples collected, a straightforward method to accelerate measurements is reducing the number of sampling points \cite{DimensionSampling}. Different strategies have been developed to minimize sampling points without sacrificing measurement accuracy \cite{optimal-sampling}, including the use of nonuniform sampling trajectories, compressive sensing, and adaptive sampling strategies. Compared to the number of sampling points at the Nyquist rate, nonuniform sampling trajectories can reduce the number by more than 50\% \cite{Spiral_pattern-eval,spiral-sampling_FF-recon,non-uniform_chen}. However, to satisfy the requirement for further processing by the fast-Fourier transform (FFT) on a uniform grid, some of the acquired samples are redundant and introduce limited additional information  \cite{optimal-sampling}. Compressive sensing (CS) is a powerful signal processing method that reduces sampling points via data sparsity, with demonstrated improvements in domains facing similar challenges such as MRI imaging \cite{compressive-sensing-mri,CompressiveSensing-review,compressiveSensing_ANN}. Recent advancements have also been made applying these methods to near-field antenna measurements \cite{compressive-sensing_NF_array,compressive-sensing-bayesian,compressive-sensing-NF-reliable,compressivesensing-bayesian-NF,minSample_CS,fieldRecon_CS}. Although CS substantially reduces measurement time, it can be computationally intensive, and it requires sufficient data sparsity and prior knowledge of the antenna under test (AUT) to be effective \cite{phaseless-faulty-diagnosis,compressive-sensing-bayesian}. Adaptive sampling, on the other hand, presents a real-time decision-making scheme to strategically collect data points based on previous samples. Despite its noteworthy achievements in reducing data acquisition time by focusing on critical regions with high variability \cite{adaptive-sampling_1,adaptive-sampling-SSAS,adaptivesampling_magazine}, some challenges to robustness and feasibility have been reported due to its heavy dependency on hyperparameters and initial sample selection \cite{batch-selection}. Other pioneering work includes utilizing spatial convolution and field similarity to reduce the total measurement time for broadband data collection \cite{Field-similarity}, and adopting phase-less sampling strategies to reduce system complexity and acquisition time \cite{Phaseless_linearized,phaseless-faulty-diagnosis,phaseless-investigation,phaseless-single-cut,phaseless-spherical}.  \par

Here, to further reduce sampling points without compromising the accuracy of antenna measurements, we draw inspiration from the recent advances in deep learning for image super-resolution and explore a deep learning-based fast NF antenna measurement workflow. Specifically, we develop a convolutional neural network model, Near-field Super-resolution Network (NFS-Net), which takes under-sampled near-field data as input and predicts high-resolution counterparts that meet the Nyquist criterion. These predicted high-resolution data are subsequently processed by the Near-Field to Far-Field (NF2FF) algorithm to obtain far-field radiation patterns. To fully leverage the ability of deep learning methods in accelerating antenna measurement, we demonstrate an end-to-end workflow and highlight three major innovative contributions:\par

\begin{enumerate}
    \item Dataset: Deep learning is a data-driven method. Therefore, we construct a dataset containing 13,672 unique near-field patterns (including magnitude and phase) for network training. The dataset is obtained through full-wave EM simulations of antennas of different sizes, operating frequencies, and directivity. The diverse antenna patterns enable effective learning of high-resolution prediction and improve generalizability of data-driven methods.
    
    \item Loss function: The loss function plays a critical role in network optimization and the ultimate effectiveness of training. For the phase prediction network, considering the periodicity of phase angles, we design a unique periodic phase loss function. This loss function significantly improves the network's ability to predict substantial phase variations in near-field data. For the magnitude prediction network, we combine the conventional mean-absolute error loss with the multi-scale structural similarity loss to drive network optimization, enhancing the network's ability to predict high-resolution magnitude. 
    
    \item The diverse dataset and specially tailored loss functions, combined with the advantages of U-Net architecture, produce a well-trained NFS-Net that can accurately reconstruct fully-sampled data from severely undersampled near-field data. Only 11\% of the sampling points are needed to fully restore the points satisfying the Nyquist rate, thereby substantially reducing measurement time while maintaining the accuracy of the final far-field antenna patterns.
\end{enumerate}

The performance of the well-trained network has been demonstrated using simulation and measurement data of antennas with different types and operating frequencies. The experimental results show the network has both accuracy and generalizability in enhancing the resolution of near-field data. This is the first work to use the capabilities of deep learning methods in addressing the challenging tasks of reconstructing near-field data in both magnitude and phase. The performance of the algorithm strongly supports a new approach for fast NF antenna measurement. \par

The rest of the paper is organized as follows. Section II describes the proposed end-to-end workflow of the fast NF antenna measurement and the NF2FF algorithm. Section III presents the proposed network design, dataset generation, loss function, and implementation details. Section IV presents the experimental results in both simulation and measurement followed by a discussion in Section V, and Section VI concludes the paper.  \par

\begin{figure*}[t]
	\centering
	\centerline{\includegraphics[width = 1\linewidth]{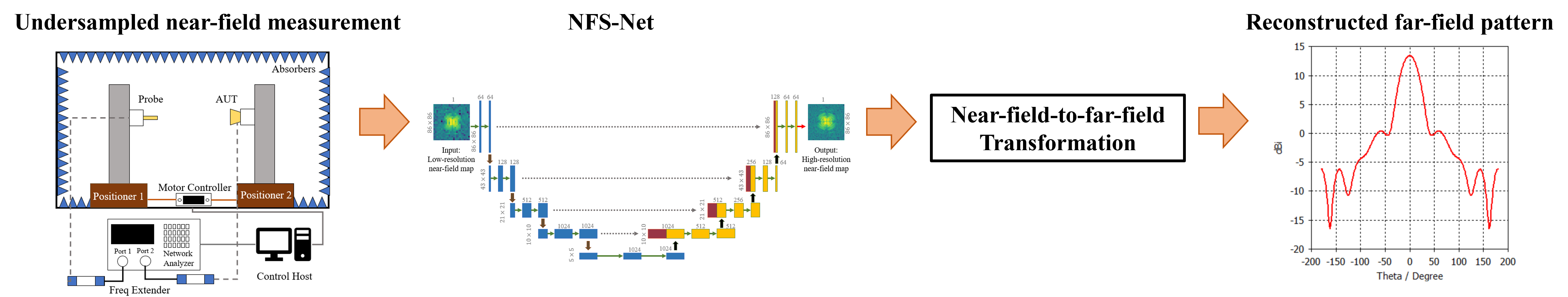}}
	\caption{The workflow of the proposed end-to-end fast NF antenna measurement. It starts with the undersampled measurement of $E_x$ and $E_y$ in the near-field of the AUT. The undersampled near-field data map is then fed into the NFS-Net to restore its fully-sampled counterpart, which is then further processed by the NF2FF algorithm to reconstruct the far-field radiation pattern.}
	\label{Fig:system_workflow}
\end{figure*}
\begin{figure*}[t]
	\centering
	\centerline{\includegraphics[width = 1\linewidth]{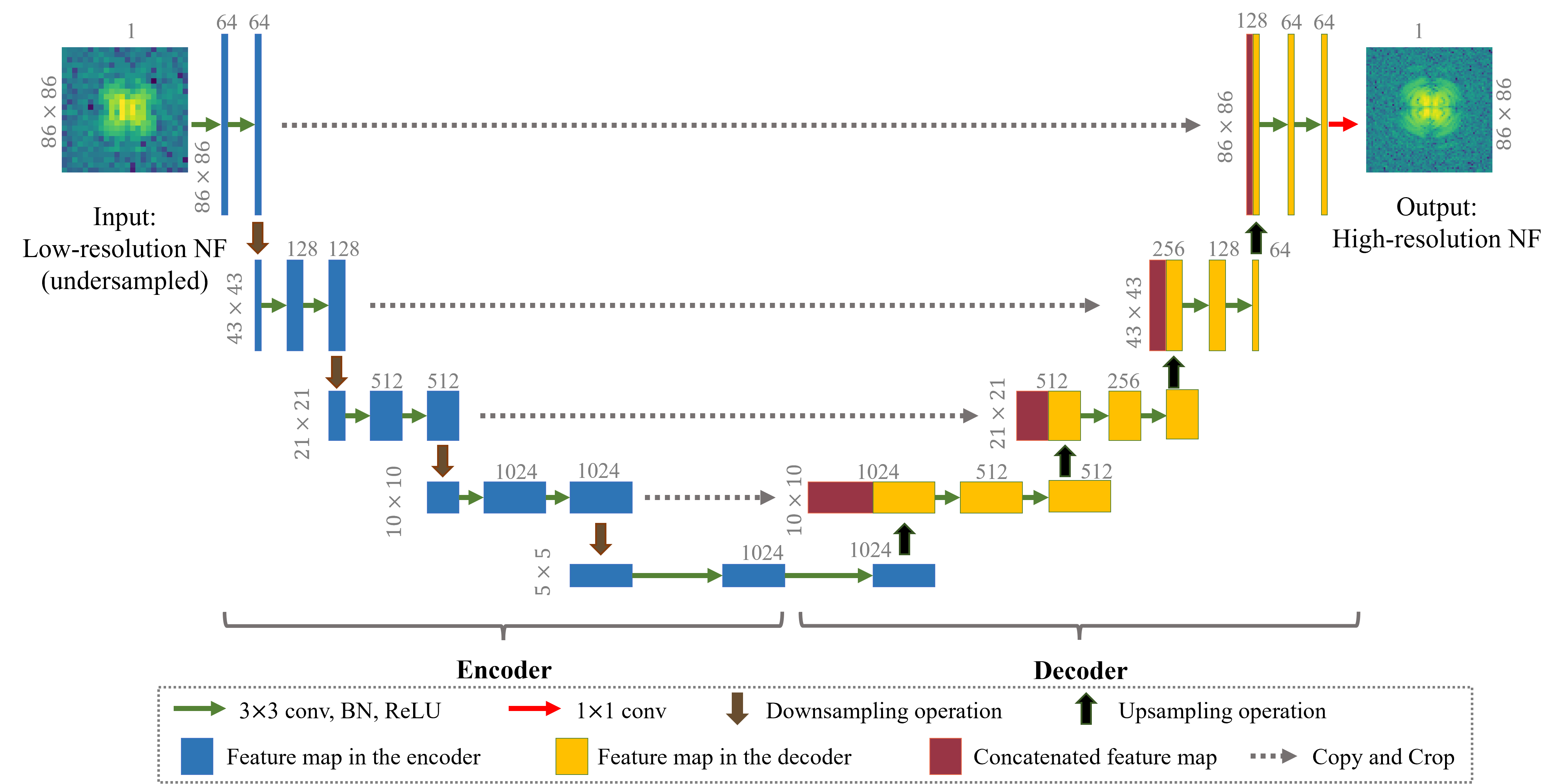}}
	\caption{The overall architecture of NFS-Net. It consists of an encoder, a decoder, and skip connections between them to preserve high-resolution features and extract informative contents. Detailed dimensions and channels for the training process can also be found in numerical annotations.}
	\label{Fig:net}
\end{figure*}

\section{Fundamentals of Near-Field-to-Far-Field Transformation and the Fast Near-Field Antenna Measurement Workflow}\label{sec2}

In this section, the analytical solutions of the NF2FF algorithms are presented in Section II-A. These algorithms are crucial for accurately transforming near-field data into far-field radiation patterns. Following this, Section II-B describes an integrated,  end-to-end system workflow that combines both the NF2FF algorithms and the Near-field Super-resolution Network (NFS-Net). This comprehensive workflow demonstrates how our proposed deep learning network and well-established analytical methods work in conjunction to enhance the efficiency and practicability of antenna characterization.\par

\subsection{NF2FF algorithms  }
The planar near-field to far-field (NF2FF) transformation relies on plane wave (modal) expansion using Fourier transform techniques. This method decomposes any monochromatic wave into a superposition of plane waves, resolving them in varying amplitudes and different directions to facilitate the conversion of near-field data into far-field radiation patterns \cite{3M-Field,balanis}. 
The NF2FF transformation starts from measuring the tangential components $E_{xa}$ and $E_{ya}$ over the scanning plane ($z'=d$) in the near field. The corresponding $x$ and $y$ components of the plane wave spectrum, ${f_{x}}\left(k_x, k_y\right)$ and ${f_{y}}\left(k_x, k_y\right)$, are determined by

\begin{equation}
\label{equ_3}
\scriptsize
f_{x}\left(k_x, k_y\right)=\int_{-b/2}^{+b/2} \int_{-a/2}^{+a/2} E_{xa}\left(x',y',z'=d\right) e^{+j\left(k_x x'+k_y y'\right)} d x' d y',
\end{equation}

\begin{equation}
\label{equ_4}
\scriptsize
f_{y}\left(k_x, k_y\right)=\int_{-b/2}^{+b/2} \int_{-a/2}^{+a/2} E_{ya}\left(x',y',z'=d\right) e^{+j\left(k_x x'+k_y y'\right)} d x' d y',
\end{equation}
where $a$ and $b$ are the width and height of the measurement plane. Then the electric field in the far field can be obtained by
\begin{equation}
\label{equ_5}
E_{\phi}(r, \theta, \varphi) \simeq j \frac{ke^{-jkr}}{2\pi r} \cos \theta \left[ -f_x \sin \varphi + f_y \cos \varphi \right]
\end{equation}

\begin{equation}
\label{equ_6_final}
E_{\theta}(r, \theta, \varphi) \simeq j \frac{ke^{-jkr}}{2\pi r} \left[ f_x \cos \varphi + f_y \sin \varphi \right]
\end{equation}
\par

From \eqref{equ_3}-\eqref{equ_6_final}, ${E}_{x a}$ and ${E}_{y a}$ are first measured at a distance $d$ from the antenna aperture, usually $3 \lambda$ to $5 \lambda$  in air, at a grid resolution of at least $0.5 \lambda$ to satisfy the Nyquist rate. The measurement area is recommended to be truncated with edges at least 40 dB lower than the max field strength within the plane \cite{IEEE_NF_recom}. 

\subsection{Proposed end-to-end workflow for fast NF antenna measurement}
 The workflow of the proposed end-to-end fast NF antenna measurement is shown in Fig. \ref{Fig:system_workflow}. It starts with a planar motion stage to traverse through physical coordinates in the near-field space of the antenna under test (AUT) to acquire undersampled $E_x$ and $E_y$ data. The undersampled near-field data is first fed into the NFS-Net to restore the fully-sampled data that satisfies the Nyquist rate. Then, the restored fully-sampled data is further processed by the NF2FF algorithm as described in Section II. A to calculate the AUT’s far-field radiation pattern.  \par

\section{Near-Field Super-Resolution Neural Network}\label{sec3}

\subsection{Dataset Preparation}
The generalizability of deep learning methods relies on the diversity and comprehensiveness of the training dataset. To ensure that NFS-Net has good generalizability for different antenna patterns, we constructed a comprehensive training and testing dataset. This dataset is built by performing full-wave electromagnetic simulations using CST Studio Suite on 24 different antenna models with operating frequencies ranging from 1 to 10 GHz. The full list of antennas is provided in Table \ref{table:antList}. Each antenna model was selected with varying parameters, including dimensions, operating frequency, polarization, and directivity, to ensure the diversity and representativeness of antennas in the dataset. The simulations follow the IEEE Standard recommendations on the best practices of near-field measurement to emulate realistic anechoic chamber characterization \cite{IEEE_NF_recom}. For each unique antenna, the simulated near-field data (magnitude and phase of the $E_x$ and $E_y$ components) are captured over a predefined spatial grid at the Nyquist rate, with a distance between $3 \lambda$ and $5 \lambda$ from the AUT. The simulation boundary is set to be up to 1.2 m to be consistent with realistic NF measurement facilities. This process results in 3418 unique near-field magnitude and phase maps. \par

\begin{table}[tb]
	\caption{List of Antennas Simulated for the Training Dataset}
	\centering
	\begin{tabular}{p{3.5cm}<{\centering} p{1cm}<{\centering} p{1.2cm}<{\centering} p{1cm}<{\centering} p{1.1cm}<{\centering}}
		\hline
		\hline
		\textbf{Antenna type}   & \textbf{Fcenter (GHz)} & \textbf{Directivity} & \textbf{Array}\\
		\hline
\textbf{Bowtie (regular)}  &  1.5    &	Omni &   /  \\
\textbf{Bowtie (complementary)}  &  10    &	Omni &   /  \\
\textbf{Bowtie (array)}  &  5    &	Directional &   Array  \\
\textbf{Dielectric resonator}  &  4.2    &	Omni &   /  \\
\textbf{Dipole (wire)}  &  9    &	Omni &   /  \\
\textbf{Dipole (cross) \cite{cross_dipole_haihan}}  &  2    &	Omni &   /  \\
\textbf{Dipole (folded)}  &  2.4    &	Omni &   /  \\
\textbf{Dipole (array)}  &  2.4    &	Directional &   Array  \\
\textbf{Helix (normal)}  &  5    &	Omni &   /  \\
\textbf{Helix (axial)}  &  2.5    &	Directional &   /  \\
\textbf{Horn (E-plane)}  &  10    &	Directional &   /  \\
\textbf{Horn (H-plane)}  &  10    &	Directional &   /  \\
\textbf{Leakywave}  &  10    &	Directional \& Omni &   /  \\
\textbf{Loop (single)}  &  2    &	Omni &   /  \\
\textbf{Loop (array)}  &  4    &	Directional &  Array  \\
\textbf{Patch (single)}  &  7    &	Omni &  /  \\
\textbf{Patch (1x4)}  &  5    &	Directional &  Array  \\
\textbf{Patch (2x2)}  &  7    &	Directional &  Array  \\
\textbf{Patch (dual-polarized)}  &  10    &	Directional \& Omni &  Array  \\
\textbf{Slot}  &  9    &	Directional &  /  \\
\textbf{Slotted waveguide}  & 10    &	Directional \& Omni&  /  \\
\textbf{Reflector}  & 10    &	Directional&  /  \\
\textbf{Vivaldi (dual-polarized) \cite{Vivaldi_haihan}}  &  1    &	Directional &   /  \\
\textbf{Vivaldi (planar)}  &  5.8    &	Directional &   /  \\

		\hline
		\hline
	\end{tabular}
	\label{table:antList}
\end{table}

The number of fully-sampled near-field maps was further augmented by rotating each map by $90^\circ$ three times, producing 13,672 fully-sampled  near-field maps. This augmentation process increases the pattern diversity in the dataset, enhancing the robustness and generalizability of the data-driven methods across different scenarios. \par

The fully-sampled near-field maps are then downsampled by a factor of 3 by uniformly selecting 1 point from every 3$\times$3 grid, producing under-sampled maps. The fully-sampled map and the corresponding under-sampled map form a data pair in the dataset. Therefore, the constructed dataset consists of 13,672 data pairs for network training and testing. \par

\subsection{Network Architecture}
Inspired by the classic U-Net for its unique, symmetric shape to preserve high-resolution feature and enhance data utilization \cite{RUNet_new,U-Net-original,DL-MRI}, we propose the NFS-Net to take advantage of such architecture and  efficiently transform undersampled near-field data into fully-sampled data. \par 

The network architecture is shown in Fig. \ref{Fig:net}. It consists of an encoder, a decoder, and skip connections. The encoder includes four stages, each with two 3×3 convolutional layers, a stride of 1, and a padding of 1, followed by batch normalization and rectified linear (ReLU) activation function. Downsampling is performed using max-pooling layers with 2×2 kernels and a stride of 2.  The decoder mirrors the encoder, using an upsampling layer with a kernel size of 2, an upscale factor of 2, and a stride of 2. This is followed by concatenation with the corresponding encoder features and further processing by two 3×3 convolutional layers. The bottleneck layer captures high-level features between the encoder and decoder. Skip connections between the encoder and decoder preserve spatial information across layers, which is crucial for maintaining output fidelity. The final 1×1 convolutional layer reduces the number of output channels to one, providing the estimated fully-sampled near-field data.  \par

\subsection{Loss Function}
The same network architecture, i.e. the NFS-Net, is separately trained to learn to produce high-resolution magnitude and phase data. To effectively train NFS-Net to extract useful features from near-field magnitude and phase data and achieve the high-resolution effect, we designed unique loss functions for magnitude and phase training.

\subsubsection{Magnitude loss function}
The magnitude loss function combines the mean absolute error (MAE) and the multi-scale structural similarity (MS-SSIM) loss, which is expressed as
\begin{equation}
\label{equ_loss_mag}
{\rm L_{mag}}=\alpha_{mag}*{\rm MAE}\left(y,\hat{y}\right)+\beta_{mag}*{\rm L_{MS\text{-SSIM}}}\left(y,\hat{y}\right),
\end{equation}
where $y$ is the predicted fully-sampled magnitude data, $\hat{y}$ is the ground truth, $\alpha_{mag}$ and $\beta_{mag}$ are weighted coefficients for magnitude to balance the contributions of MAE and MS-SSIM loss. The MAE term computes numerical differences between predicted fully-sampled magnitude data $y$ and ground truth $\hat{y}$, while MS-SSIM captures the perceptual quality difference between the two \cite{MS-SSIM}. \par

MAE is calculated as
\begin{equation}
\label{equ_mae}
{\rm MAE}\left(y,\hat{y}\right)=\frac{1}{H\times W}\sum_{i,j}\left|y_{i,j}-{\hat{y}}_{i,j}\right|,
\end{equation}
where $H \times W$ is the dimension of the image, and $i$ and $j$ are the indices for pixels. \par
MS-SSIM measures the visual differences between the predicted fully-sampled magnitude data $y$ and ground truth $\hat{y}$ in terms of the luminance $l\left(y,\hat{y}\right)$, contrast $\ c\left(y,\hat{y}\right)$ and structure $s\left(y,\hat{y}\right)$, which is calculated as 
\begin{equation}
\label{equ_msssim}
{\rm MS\text{-SSIM}}\left(y,\hat{y}\right)=\left[l_M\left(y,\hat{y}\right)\right]^{\alpha_M}\cdot\prod_{k=1}^{M}{\left[c_k\left(y,\hat{y}\right)\right]^{\beta_k}\left[s_k\left(y,\hat{y}\right)\right]^{\gamma_k}}.
\end{equation}
In (\ref{equ_msssim}), $M$ represents the number of scales the image is analyzed, $k$ represents the current scale level in the calculation, $\alpha_M$, $\beta_k$ and $\gamma_k$ are the coefficients to adjust weights in different scales. The luminance $l\left(y,\hat{y}\right)$, contrast $\ c\left(y,\hat{y}\right)$ and structure $s\left(y,\hat{y}\right)$ are given by

\begin{equation}
\label{equ_l}
l\left(y,\hat{y}\right)=\frac{2\mu_y\mu_{\hat{y}}+C_1}{\mu_y^2+\mu_{\hat{y}}^2+C_1},
\end{equation}
\begin{equation}
\label{equ_c}
c\left(y,\hat{y}\right)=\frac{2\sigma_y\sigma_{\hat{y}}+C_2}{{\sigma_y}^2+{\sigma_{\hat{y}}}^2+C_2},
\end{equation}
\begin{equation}
\label{equ_s}
s\left(y,\hat{y}\right)=\frac{\sigma_{y\hat{y}}+C_3}{\sigma_y\sigma_{\hat{y}}+C_3},
\end{equation}
where $\mu_y$ and $\mu_{\hat{y}}$ are the means of $y$ and $\hat{y}$, $\sigma_y$ and $\sigma_{\hat{y}}$ are the variances of $y$ and $\hat{y}$, $\sigma_{y\hat{y}}$ is the covariance of $y$ and $\hat{y}$. $C_1$, $C_2$, and $C_3$ are small constants to stabilize the division\cite{MS-SSIM}. The loss function for MS-SSIM is defined as 

\begin{equation}
\label{equ_loss-ssim}
{\rm L_{MS\text{-SSIM}}}\left(y,\hat{y}\right)=1-{\rm MS\text{-SSIM}}\left(y,\hat{y}\right).
\end{equation}

\subsubsection{Phase loss function}

 Since the acquired field data must be normalized to values between 0 and 1 to facilitate convergence during training on PyTorch tensors \cite{pytorch}, wrapped phase angles are used for phase representation to avoid the extensive range of values presented by unwrapped phase angles. However, phase wrapping introduces challenges in defining the loss function.  When two phase angles are close to the wrapping point, for example, $-179^\circ$ and $0^\circ$, they appear significantly different when evaluated by the conventional MAE loss function. However,  the difference between the two is actually small in the context of the phase angle. Therefore, to address the issue caused by phase wrapping, we designed a novel loss function, namely Periodic Phase Loss ($L_{pp}$), which compensates for the periodic nature of phase angle during network training. The $L_{pp}$ is defined as

\begin{equation}
\label{equ_lpp}
\mathrm{L_{pp}}(z, \hat{z}) = \frac{1}{H \times W} \sum_{i,j} \min \left( \left| z_{i,j} - \hat{z}_{i,j} \right|, \left| z_{i,j} - \hat{z}_{i,j} - 1 \right| \right),
\end{equation}
where $z$ is the predicted fully-sampled phase data, $\hat{z}$ is the ground truth, both are normalized to [0,1]. \par
Similar to the loss function for magnitude, the final loss function for phase also combines the  MS-SSIM loss with the newly defined periodic phase loss. The final phase loss function is defined as
\begin{equation}
\label{equ_loss_phase}
{\rm L_{phase}}=\alpha_{phase}*{\rm L_{pp}}\left(z,\hat{z}\right)+\beta_{phase}*{\rm L_{MS\text{-SSIM}}}\left(z,\hat{z}\right),
\end{equation}
where $\alpha_{phase}$ and $\beta_{phase}$ are weighted coefficients for phase training to balance the effect of corresponding losses.

\begin{figure*}[t]
	\centering
	\centerline{\includegraphics[width = 1\linewidth]{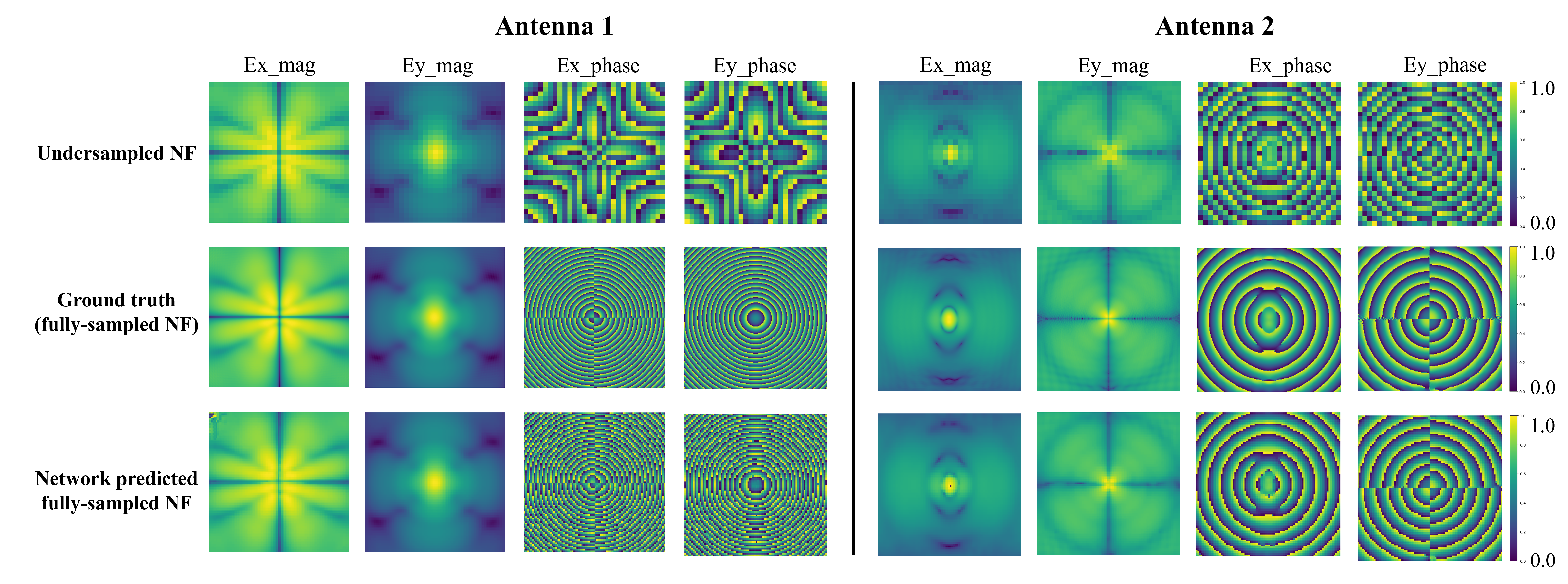}}
	\caption{Comparison of the network's capability to conduct near-field super-resolution on data with a downsampling factor of 3. Undersampled NF and ground truth represent the low-resolution input and the fully-sampled, high resolution output of the network, with predicted NF showing the network's estimations based on low-resolution data. Antenna 1 is a corrugated horn antenna working at 8.7 GHz, while antenna 2 operates at 3 GHz as a Yagi antenna. The range of values is [0, 1] as shown on the color bar. }
	\label{Fig:field_comp}
\end{figure*}
\subsection{Implementation details}
The 13,672 pairs of data in the dataset are randomly split into 80\% training data and 20\% testing data. All the near-field maps are resized to $86\times86$ and normalized to a range of [0, 1] for training. The input data size is chosen to accommodate our computing resource limitations and training time. However, it is important to note that the U-Net structure is highly versatile and can accept 2D matrices of different sizes, making it suitable for other sampling surface sizes as well.  \par
The network is implemented in PyTorch and trained on a single NVIDIA 4090 GPU (24-GB GPU memory). The training process is based on the loss functions as described in Section III. C. To better understand the impact of hyperparameter choices on the training outcomes, we conducted a comprehensive evaluation of the training results across three major hyperparameters: batch size, learning rate, and the number of network layers, with the comparison shown in Table \ref{table:comparenetwork}. The batch size in network training refers to the number of samples processed in one pass through the network during each training step. A large batch size can accelerate training speed and reduce noisy gradients, whereas a small batch size can lower the demand for computing resources and improve generalizability. The learning rate is another important parameter that significantly impacts training outcomes, as it controls the size of the steps taken when updating the model’s weights. A high learning rate can speed up training by taking larger steps, whereas a low learning rate results in more precise steps and better training stability. The number of network layers also plays a critical role in training, as features are extracted from the input data through these layers. A network with fewer layers consumes fewer computing resources, whereas a deeper network can extract comprehensive features more accurately.\par

After careful testing and exploration, balancing training time, training quality and computing resources, we found that a network of 64 layers and mini-batch size of 15, with $\alpha_{mag}$ = 1, and $\beta_{mag}$ = 1 for magnitude loss function, and $\alpha_{phase}$ = 0.6, and $\beta_{phase}$ = 0.4 for phase loss function, provides the best training outcomes. The network is optimized using the ADAM optimizer \cite{adam}. The initial learning rate is set to 0.001, and it decreases by a factor of 10 every 50 and 75 epochs for magnitude and phase training, respectively, resulting in total training epochs of 200 and 300. The total training time is approximately 4 hours for the magnitude network and 6 hours for the phase network. The progression of training and validation loss decay is shown in Fig. \ref{Fig:loss decay_side_by_side}. The effectiveness of the neural network training is clearly demonstrated, and the saturation observed toward the end of the process confirms that the stopping point is adequate to ensure high training quality.  \par

\section{Experimental results}\label{sec4}
The performance of the well-trained NSF-Net has been examined using both simulated and measured data. In this section, we first present near-field super-resolution examples on 3 simulated antennas, followed by NF2FF reconstruction based on these fully-sampled near-field data predicted by the network. Then we further explore the network's capability using measurement data acquired on different antennas working at frequencies significantly higher than those in the training datasets. The results demonstrate the effectiveness of the proposed network on ideal full-wave EM simulation data and on actual measurement results, showcasing its substantial generalizability across scenarios. 
The undersampled near-field data, using only 11\% of the samples required at Nyquist rate, can achieve far-field patterns comparable to fully-sampled data through neural network-based super-resolution reconstruction. This significant reduction in sampling points greatly accelerates NF antenna characterization without compromising its performance.
\par

\begin{figure}[t]
	\centering
	{\includegraphics[width = 0.9\linewidth]{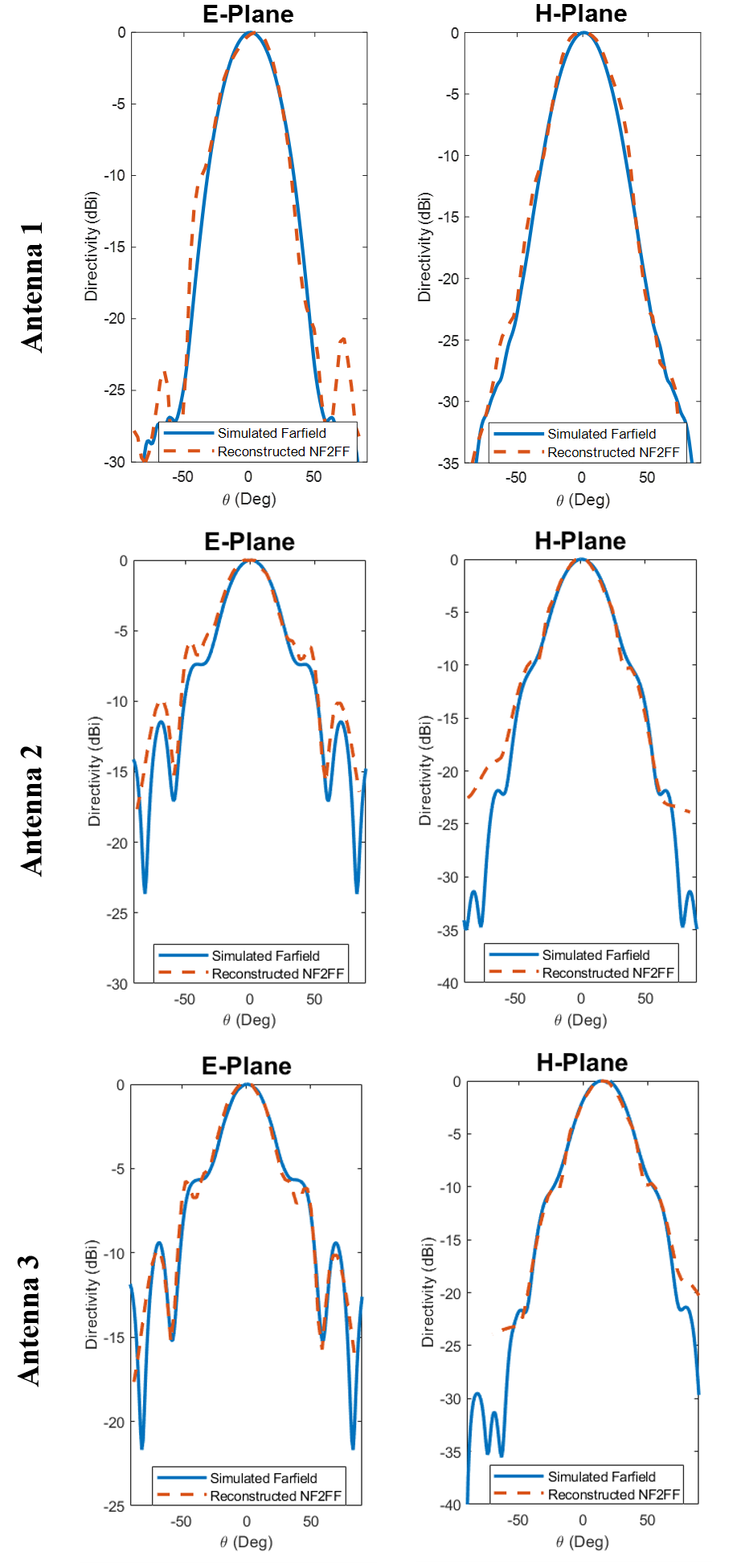}}
	\caption{Comparison of the simulated (fully-sampled) and reconstructed (3$\times$ downsampled) far-field patterns in both E and H plane. Antenna 1 is an X-band corrugated horn antenna, Antenna 2 is a Yagi antenna operating at 3 GHz, and Antenna 3 is a modified version of Antenna 2 tilted at an angle of $15 ^\circ$. None of these types of antennas is included in the training dataset.}
	\label{Fig:sim_result_recon}
\end{figure}

\subsection{Results on simulation data}

The network's super-resolution performance for the amplitude and phase data of the $x$ and $y$ components of the electric field is validated on two different antennas. Antenna 1 is an X-band corrugated horn antenna working at 8.7 GHz, Antenna 2 is a Yagi antenna operating at 3 GHz, and Antenna 3 is a modified version of Antenna 2 tilted at an angle of 15°. None of these types of antennas are included in the training dataset. The results are shown in Fig. \ref{Fig:field_comp}.  In the figure, the undersampled NF and network prediction represent the undersampled near-field input to the network and the fully-sampled output data from the network, with values normalized between 0 and 1. The ground-truth data is sampled at Nyquist rate with a sampling resolution of $\lambda/2$, and the undersampled NF data is obtained by undersampling the ground-truth data by a factor of 3, resulting in a sampling resolution of $3\lambda/2$. The comparison results shown in Fig. \ref{Fig:field_comp} present that the fully-sampled amplitude and phase data predicted by the network closely resemble the ground truth. The errors between the network's predicted data and the ground truth are listed in Table  \ref{table:comparison_loss_decay}. The low MAE of the magnitude, the low $L_{pp}$ of the phase, and the high MS-SSIM of both the magnitude and phase in the numerical comparison, combined with the visualized results shown in Fig. \ref{Fig:field_comp}, confirm the network's outstanding super-resolution capability. \par

\begin{figure}[tb]
	\begin{flushleft}
		\begin{tabular}{@{}c@{}c@{}}
			\includegraphics[width=0.48\linewidth]{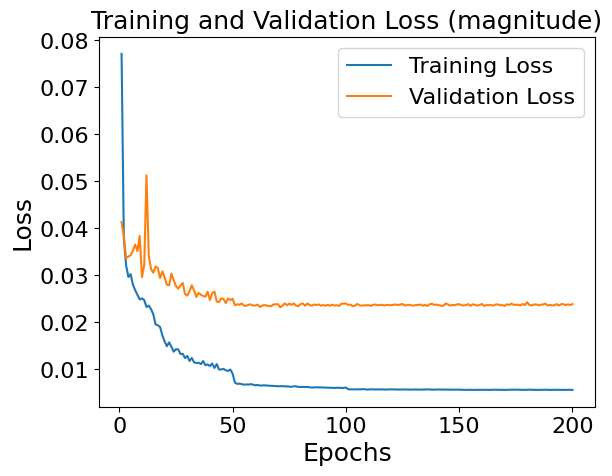} \hfill&
			\includegraphics[width=0.48\linewidth]{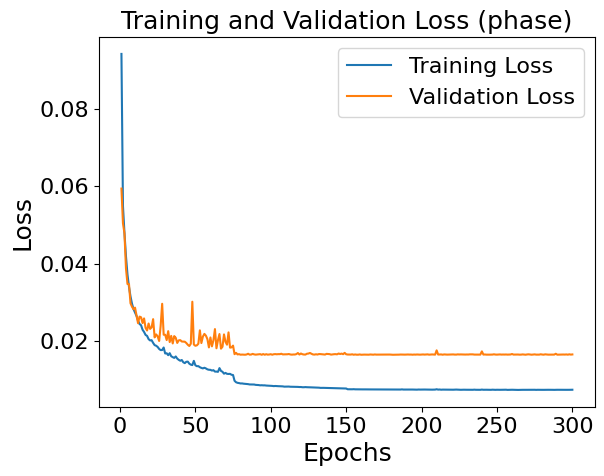} \hfill\\
			\footnotesize{(a)} & \footnotesize{(b)} \\
		\end{tabular}
	\end{flushleft}
	\caption{{Training and validation loss decay of magnitude (a) and phase (b) for the network.}}
	\label{Fig:loss decay_side_by_side}
\end{figure}

\begin{table}[tb]
	\caption{{The Impact of Hyperparameters on the Performance of the Proposed Network (bs=batch size, lr= learning rate(initial), nl=number of network layers)}  }
	\centering
	\begin{tabular}{p{3.3cm}<{\centering} p{1.5cm}<{\centering} p{1.4cm}<{\centering}}
		\hline
		\hline
		\textbf{Parameters}   & \textbf{Loss (Phase↓)} & \textbf{Loss (Mag↓)} \\
		\hline
\textbf{BS=5, LR=0.001, NL=64}  &  0.0207    &	0.0126    \\
\textbf{BS=25, LR=0.001, NL=64}  &  0.0171    &	 0.0115   \\
\textbf{BS=15, LR=0.01, NL=64}  &  0.0180    &	0.0138    \\
\textbf{BS=15, LR=0.0001, NL=64}  &  0.0236    & 0.0122  \\
\textbf{BS=15, LR=0.001, NL=32}  &  0.0209    &	0.0123     \\
\textbf{BS=15, LR=0.001, NL=128}  &  0.0153    &	 0.0119  \\
\textbf{BS=15, LR=0.001, NL=64 (proposed network)}  &   \textbf{0.0165}   &	 \textbf{0.0120}    \\

		\hline
		\hline
	\end{tabular}
	\label{table:comparenetwork}
\end{table}

After completing the first step of near-field super-resolution, the fully-sampled magnitude and phase data predicted by the network are fed into the NF2FF algorithm to reconstruct the far-field radiation patterns of the antennas. The reconstruction results (in red) are compared to the simulated far-field patterns (in blue) in Fig. \ref{Fig:sim_result_recon}. As observed in this comparison, they show remarkable agreement in both E-plane and H-plane patterns. \par

To further validate the network's generalization capability, the network's performance is tested on a modified Yagi antenna (Antenna 3) with patterns tilted at 15°. The results are also provided in Fig. \ref{Fig:sim_result_recon}. The 15° tilt is clearly visible In the reconstructed H-plane pattern, and the predicted E- and H-plane patterns closely match the simulated far-field patterns, confirming the network's ability to accurately restore fully-sampled near-field data for antennas with different patterns.\par

\begin{figure}[t]
    \begin{center}
        \begin{tabular}{@{}m{0.5cm}@{}c@{\hspace{0.5cm}}c@{}}
            \rotatebox{90}{\fontsize{10}{16}\selectfont\hspace{4.7cm}\textbf{Antenna 4}} & 
            \includegraphics[width=0.4\linewidth]{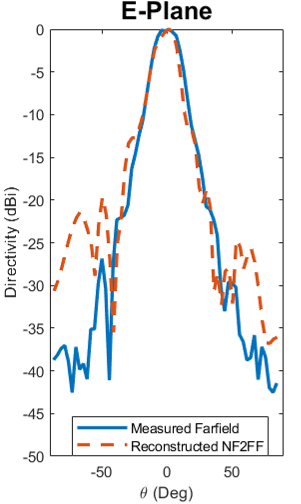} & 
            \includegraphics[width=0.4\linewidth]{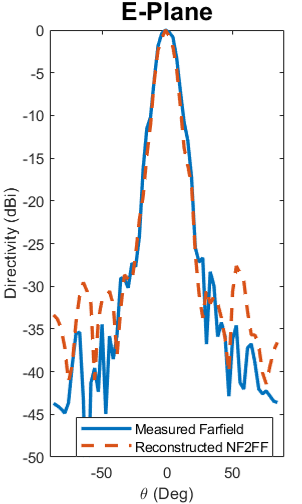} \\
            \noalign{\vskip -3cm}
            \rotatebox{90}{\fontsize{10}{16}\selectfont\hspace{4.7cm}\textbf{Antenna 5}} & 
            \includegraphics[width=0.4\linewidth]{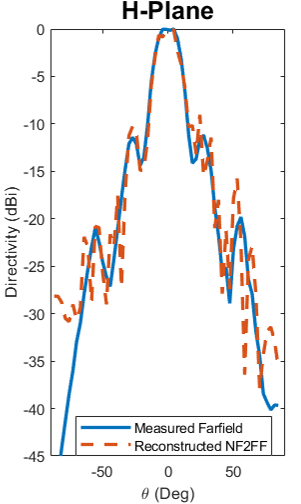} & 
            \includegraphics[width=0.4\linewidth]{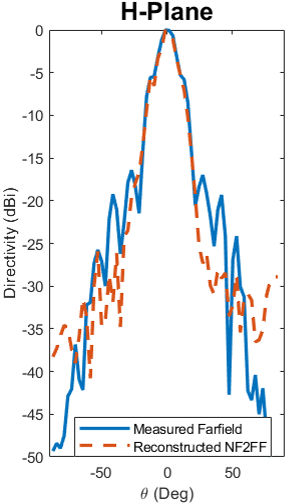} \\
            \noalign{\vskip -3.2cm}

        \end{tabular}
    \end{center}
    \caption{Comparison of the measured (fully-sampled) and reconstructed (3$\times$ downsampled) far-field patterns in both E and H plane. Antenna 4 and 5 are two horn antennas with an operating frequency at 75 GHz and 90 GHz.}
    \label{Fig:meas_result}
\end{figure}

\begin{figure}[t]
    \centering
	\centerline{\includegraphics[width = 1\linewidth]{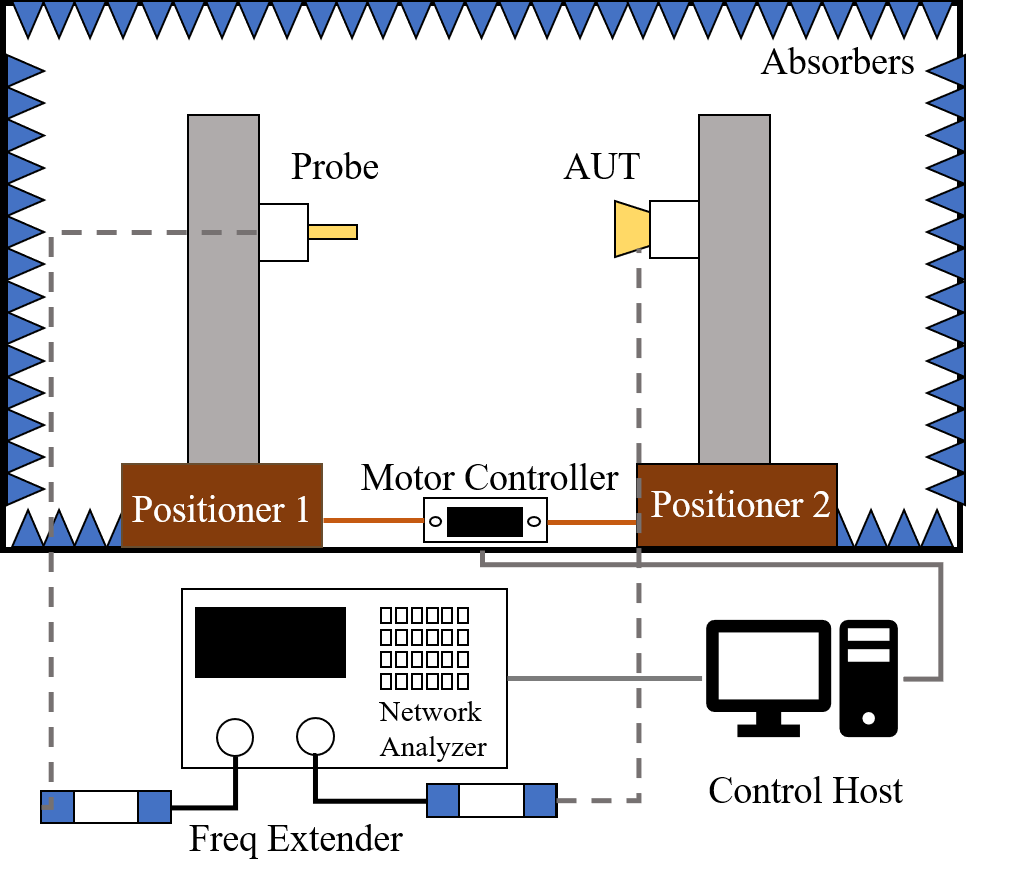}}
	\caption{The NF measurement setup used in this experiment. The system consists of a motor-controlled motion stage hosted on a vibration-absorbing optical table surrounded by absorbers. The characterization probe and the AUT are connected by a 2-port VNA and frequency extenders to acquire near-field data.}
	\label{Fig:meas_setup}
\end{figure}

\begin{figure*}[t]
	\centering
	{\includegraphics[width = 1\linewidth]{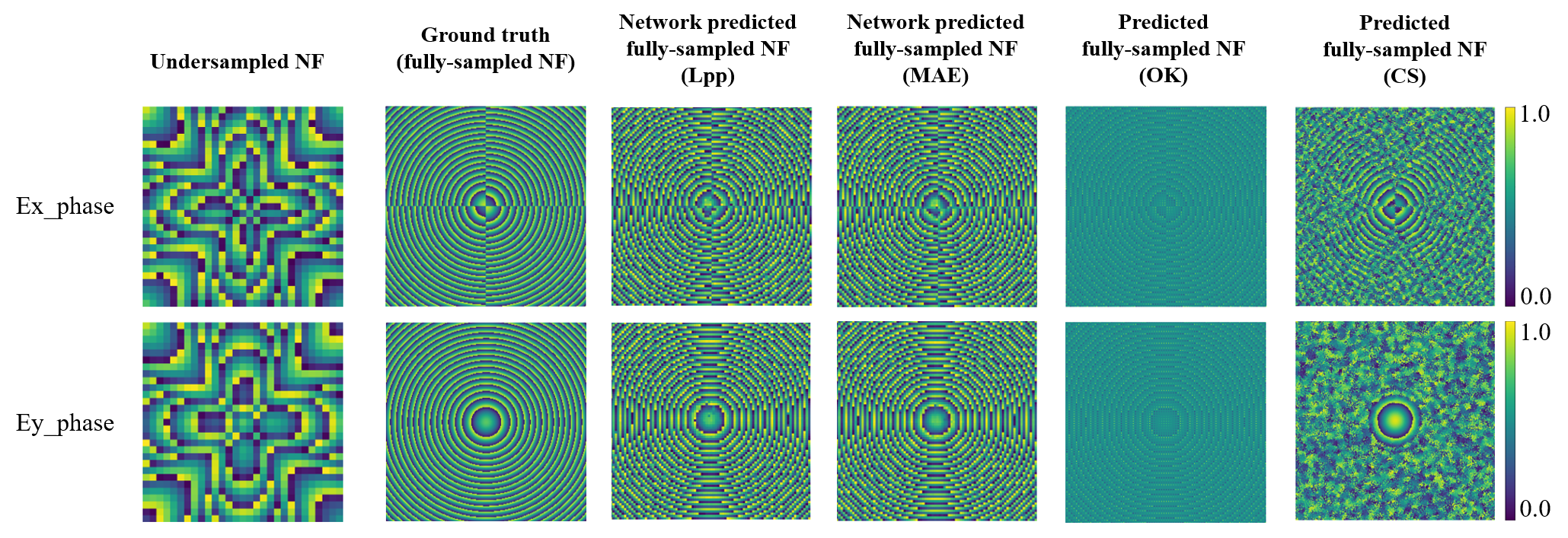}}
	\caption{Comparison of the restored near-field phase maps of Antenna 1 using different super-resolution methods and using NFS-Net trained with different loss functions. The fully-sampled NF map is refered to as ground truth. $L_{pp}$ and MAE represent periodic phase loss and mean absolute error loss that used to train the NFS-Net. OK and CS represent ordinary Kriging and compressive sensing methods.}
	\label{Fig:LPP_MAE_interp}
\end{figure*}

\begin{figure}[t]
    \begin{center}
        \begin{tabular}{cc}
            \includegraphics[width=0.38\linewidth]{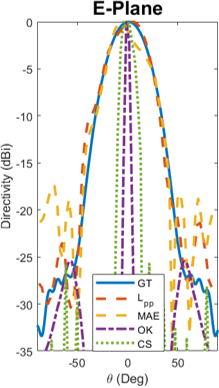} & 
            \hspace{0.3cm}
            \includegraphics[width=0.38\linewidth]{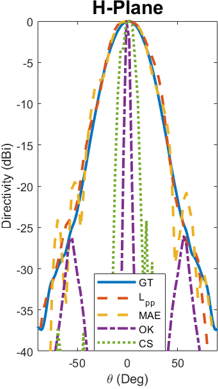} \\

        \end{tabular}
    \end{center}
    \caption{{Comparison of reconstructed far-field patterns of Antenna 1 using different super-resolution methods and using NFS-Net trained with different loss functions. The fully-sampled NF map is refered to as ground truth. $L_{pp}$ and MAE represent periodic phase loss and mean absolute error loss that used to train the NFS-Net. OK and CS represent ordinary Kriging and compressive sensing methods.}}
    \label{Fig:LPP_MAE_interp_FF}
\end{figure}

\begin{figure}[t]
	\centering
	{\includegraphics[width = 0.9\linewidth]{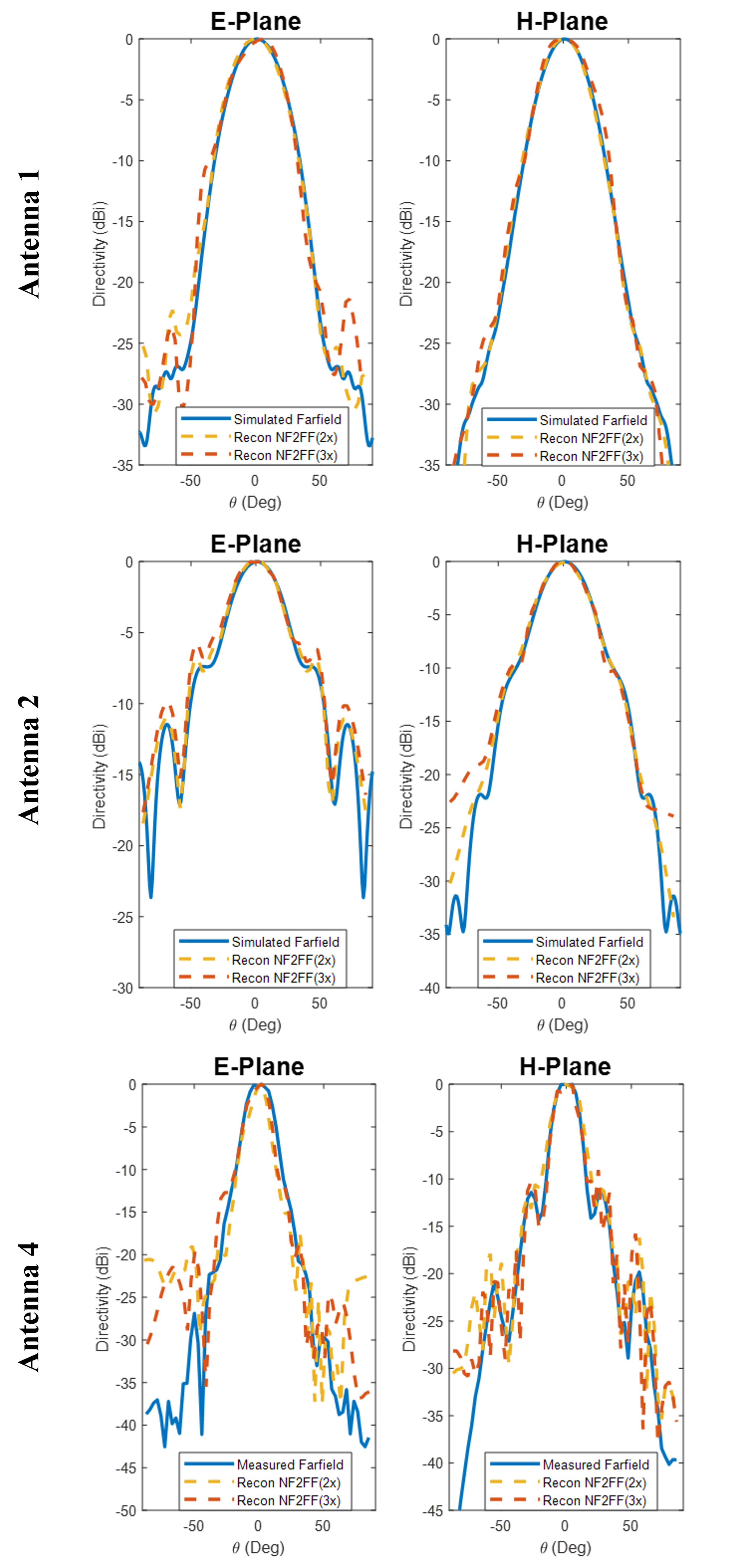}}
	\caption{{Downsamling factor analysis: comparison of the ground truth (fully-sampled) and reconstructed (3$\times$ and 2$\times$ downsampled) far-field patterns in both E and H plane. }}
	\label{Fig:DSFactor_comp_vertical}
\end{figure}

\begin{figure}[t]
	\centering
	{\includegraphics[width = 0.9\linewidth]{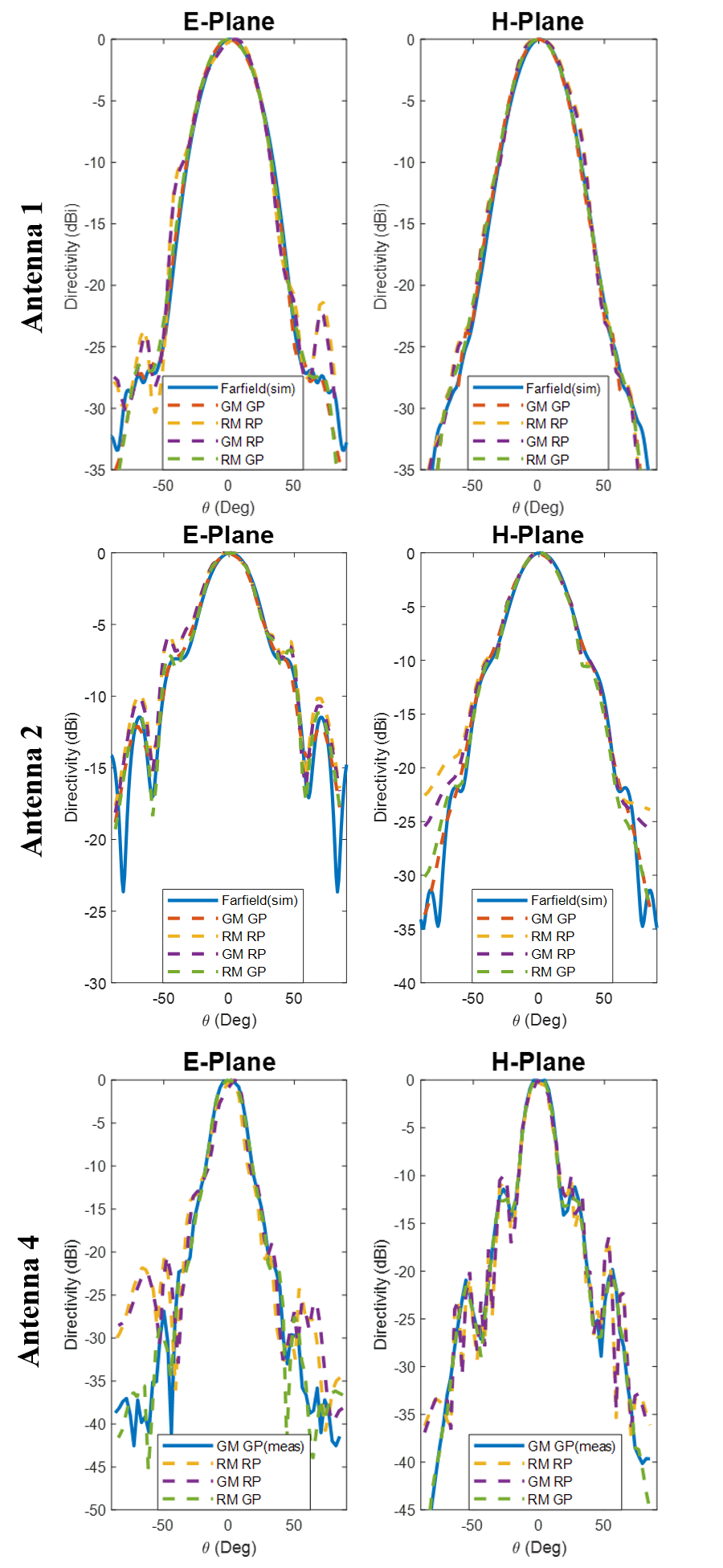}}
	\caption{{Comparison of reconstructed far-field patterns by different combinations of ground-truth (fully-sampled) and restored magnitude and phase. G = ground truth, R = restored, M = magnitude, P = phase. For example, GM RP = ground truth magnitude and restored phase.} }
	\label{Fig:GT_comp_vertical}
\end{figure}

\subsection{Results on measurement data }

The performance of the proposed model is further validated using measured NF data. The illustration of the NF measurement is shown in Fig. \ref{Fig:meas_setup}, and the detailed measurement setup is described in \cite{ManualIsabel}.  The testing probe (rectangular open-ended waveguide) and the antenna under test (AUT) are both fixed on positioners controlled by a motion controller and host computer. A vector network analyzer (R\&S ZVA-67) is used as a signal generator and receiver to characterize the near-field data, with frequency extenders (R\&S ZVA-Z110) to connect with the two testing ports. The entire measurement setup is supported on a precision optical table to minimize the effect of vibration. High frequency absorbers are also used around the setup to further reduce scatterings and reflections in the environment.\par

Two antennas are tested in the experiment. Antenna 4 is a horn antenna operating at 75 GHz with a nominal gain of 20 dBi, and Antenna 5 is a horn antenna operating at 90 GHz with a nominal gain of 24 dBi. The results are shown in Fig. \ref{Fig:meas_result}. Similar to the simulation examples, the solid blue line represents the far-field pattern obtained from fully sampled measurements at the Nyquist rate, while the dashed red line represents the far-field pattern reconstructed from 3$\times$ undersampled near-field data after network prediction. These patterns are in appropriate agreement with each other. It's important to note that the two measured antennas operate at much higher frequencies compared to the antennas in the training dataset, yet the network maintains consistency in predicting the fully-sampled near-field data. The measurement results further validate the strong generalizability of the network. \par

\subsection{Comparison of validation loss on training and experimental data}
{Since the experiments use field information collected from both simulation and measurement that is independent from the training data, we further investigated the errors occurred in magnitude and phase restoration using the proposed network for all the 5 antennas. They are compared against the values benchmarked on the training data from Fig. \ref{Fig:loss decay_side_by_side} in Table \ref{table:comparison_loss_decay}. As shown in the table, Antenna 1-3 exhibit error scales similar to those of the training data, largely due to the more ideal input data acquired through full-wave EM simulations. Antennas 4-5 show higher loss margins due to the difference between the realistic antenna measurement environment and the ideal simulation environment.\par}

\section{Discussion}\label{sec4c}
In this section, we discuss the impact of different loss functions on network's performance, compare the network's performance with other methods for accelerating NF antenna measurements, examine the effects of downsampling rates on far-field reconstruction accuracy, and analyze the source of reconstruction errors produced by the network and measurement environment. \par

\subsection{Comparison of reconstruction strategies}
{We first compare the effects of phase training using Periodic Phase Loss ($L_{pp}$) combined with $L_{MS-SSIM}$ as the loss function against the traditional MAE combined with $L_{MS-SSIM}$, while keeping magnitude and other settings the same. The comparison of the restored near-field maps is shown in Fig. \ref{Fig:LPP_MAE_interp}, and the comparison of the NF2FF reconstructed far-field patterns is shown in Fig. \ref{Fig:LPP_MAE_interp_FF}. The two cases are labeled as $L_{pp}$ and MAE, respectively. Fig. \ref{Fig:LPP_MAE_interp} shows that networks trained with MAE, which does not account for the unique nature of phase wrapping, have limited accuracy in predicting high-resolution phase map, hence it cannot accurately restore the far-field pattern as shown in Fig. \ref{Fig:LPP_MAE_interp_FF}. However, our proposed $L_{pp}$ successfully addresses this issue,  driving the network to effectively learn phase variations and achieve precise prediction of the high-resolution near-field phase map, thus resulting in more accurate far-field pattern.\par}

The reconstruction accuracy of the NFS-Net is also compared with other commonly used super-resolution methods, including the ordinary Kriging \cite{emc_kriging} and compressive sensing \cite{fieldRecon_CS}, on the downsampled near-field data at 11\% of the Nyquist rate. Fig. \ref{Fig:LPP_MAE_interp} shows the comparison of their restored phase maps and Fig. \ref{Fig:LPP_MAE_interp_FF} shows the farfield reconstruction results. The Kriging and compressive sensing methods fail to restore the fully-sampled NF phase map due to significant phase information loss in the undersampled data. In contrast, the NFS-Net, by learning from diverse NF data and being optimized using losses that include the numerical error ($L_{pp}$) and the perceptual error ($L_{MS-SSIM}$), demonstrates much higher prediction accuracy for high-resolution NF phase map. Consequently, as shown in Fig. \ref{Fig:LPP_MAE_interp_FF}, the NFS-Net achieves significantly higher accuracy in the reconstructed far-field pattern compared to the other two methods.\par

\subsection{Downsampling rate}
{To explore the effect of different downsampling rates on reconstruction quality, we compare the far-field reconstruction results with a downsampling factor of 2 (25\% of fully-sampled data, interval of 1 $\lambda$) to those with a downsampling factor of 3 (11\% of fully-sampled data, interval of 1.5 $\lambda$) for antennas 1, 2 and 4. As shown in Fig. \ref{Fig:DSFactor_comp_vertical}, a downsampling factor of 2 shows higher reconstruction accuracy. The comparison provides an intuitive guideline for practical applications: a downsampling rate of 2 is preferable when prioritizing accuracy, whereas a downsampling rate of 3 offers advantages in terms of time efficiency. \par}

\begin{table}[tb]
	\caption{{Comparison of the Validation Loss on Training and Experimental Data}}  
	\centering
	\begin{tabular}{p{2.5cm}<{\centering} p{2cm}<{\centering} p{2cm}<{\centering}}
		\hline
		\hline
		\textbf{Object}   & \textbf{Magnitude} & \textbf{Phase}\\
		\hline
\textbf{Antenna 1}  &  0.0040    & 0.0331\\
\textbf{Antenna 2}  &  0.0136    & 0.0144\\
\textbf{Antenna 3}  &  0.0130    & 0.0147  \\  
\textbf{Antenna 4}  &  0.1349    & 0.1895    \\
\textbf{Antenna 5}  &  0.1496    & 0.2178  \\
\textbf{Proposed network (training datasets)} & \textbf{0.0239} & \textbf{0.0165} \\

		\hline
		\hline
	\end{tabular}
	\label{table:comparison_loss_decay}
\end{table}

\subsection{Error contribution}
{We have conducted further analysis of the sources of error in the reconstructed far-field radiation patterns. There are two possible sources of error: inaccuracies in the restored high-resolution magnitude or inaccuracies in the restored high-resolution phase. To determine the primary source, we used different combinations of ground-truth and restored magnitude and phase to reconstruct the far-field patterns for antennas 1, 2, 4, and compared the accuracy of the reconstructed patterns. The results are shown in Fig. \ref{Fig:GT_comp_vertical}. For clarity, G, R, M, and P represent the ground truth, restored data, magnitude, and phase, respectively. For example, the combination “ground-truth magnitude and restored phase” is denoted as “GM RP”. The comparison shows that the combination “GM RP” and “RM RP” exhibit similarly large errors. Therefore, it can be concluded that the primary source of error is the inaccuracies in the restored phase. 
\par}
{Due to the periodic nature of phase, the phase information is the most challenging for the network to learn. Further improvements in phase accuracy can be achieved through advanced network architectures and more comprehensive training datasets, which are the focus of our future work.
\par}

\subsection{Measurement noise}

\begin{figure}[t]
	\centering
	{\includegraphics[width = 0.9\linewidth]{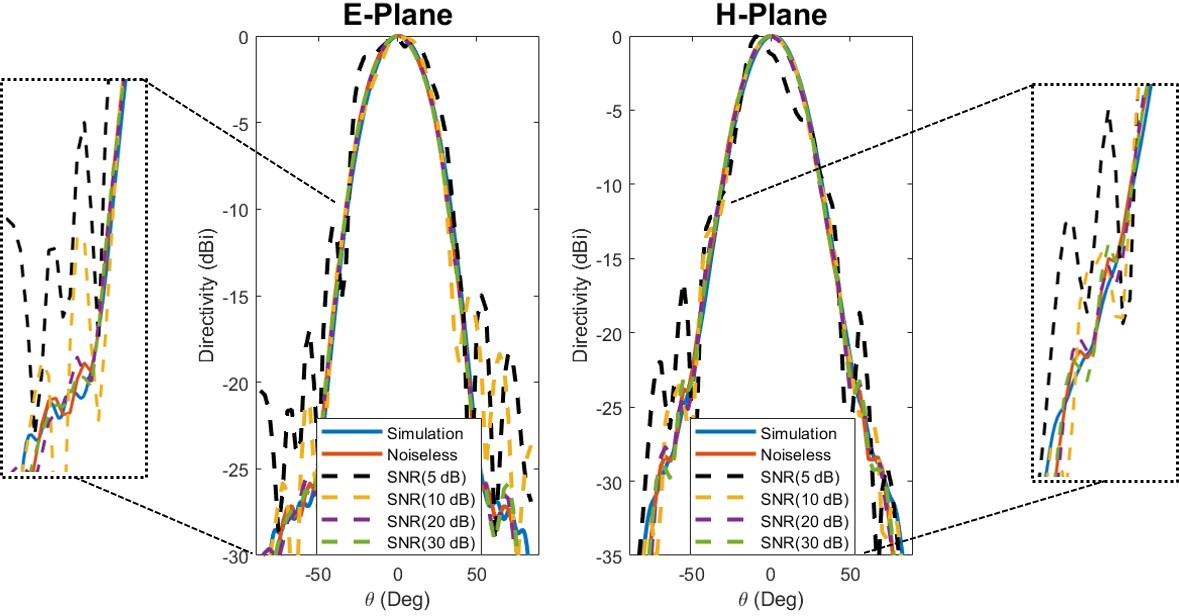}}
	\caption{{Comparison of measurement noise's impact on reconstruction quality of Antenna 1. Gaussian noise in various signal-to-noise ratios (SNR) is added to represent different noise levels in a realistic environment.}} 
	\label{Fig:Measurement SNR}
\end{figure}

{To better understand the impact of environment to reconstruction quality, we have also conducted further analysis on measurement noise. This is implemented by adding Gaussian noise to the simulation dataset of Antenna 1 at different levels of signal-to-noise ratio (SNR) to emulate the random noises in the measurement environment. As shown in Fig. \ref{Fig:Measurement SNR}, the impact on reconstruction quality is noticeable when SNR is below 20 dB, which also explains the larger error margins observed in Antenna 4-5 in Table \ref{table:comparison_loss_decay}, possibly due to noises in realistic environment.  \par}

\section{Conclusion}\label{sec5}

 In this paper, we present a  novel fast NF measurement workflow based on deep neural networks. We develop a super-resolution network, the NFS-Net, to transform significantly undersampled near-field data to fully-sampled near-field data. Subsequently, the transformed data is processed by the NF2FF algorithm to generate far-field antenna radiation patterns. To effectively train the network, we constructed a large-scale simulation dataset containing diverse near-field data and developed novel phase and magnitude loss functions. Extensive experiments on simulated and measured data have demonstrated the network's accuracy and generalizability in predicting high-resolution fully-sampled near-field data. The proposed NFS-Net and NF2FF integrated antenna measurement workflow only requires 11\% of fully-sampled data to accurately reconstruct far-field radiation patterns. The significantly reduced sampling points greatly accelerate antenna measurements. Though the method is validated using a planar NF measurement setup, it can be readily adapted to other NF platforms such as spherical and cylindrical NF measurement systems to enhance the NF measurement efficiency. \par

\section*{Acknowledgment}

The authors would like to thank Isabel Jurado Pérez at Universidad Politécnica de Cataluña, Barcelona, Spain for her contribution of near-field measurement data. This work is also supported by IEEE Antennas and Propagation Society Fellowship (APSF).

\bibliographystyle{IEEEtran}
\bibliography{IEEEabrv,Manuscript_Latex/AI_assisted_NF2FF_v2}

\begin{IEEEbiography}[{\includegraphics[width=1in,height=1.25in,clip,keepaspectratio]{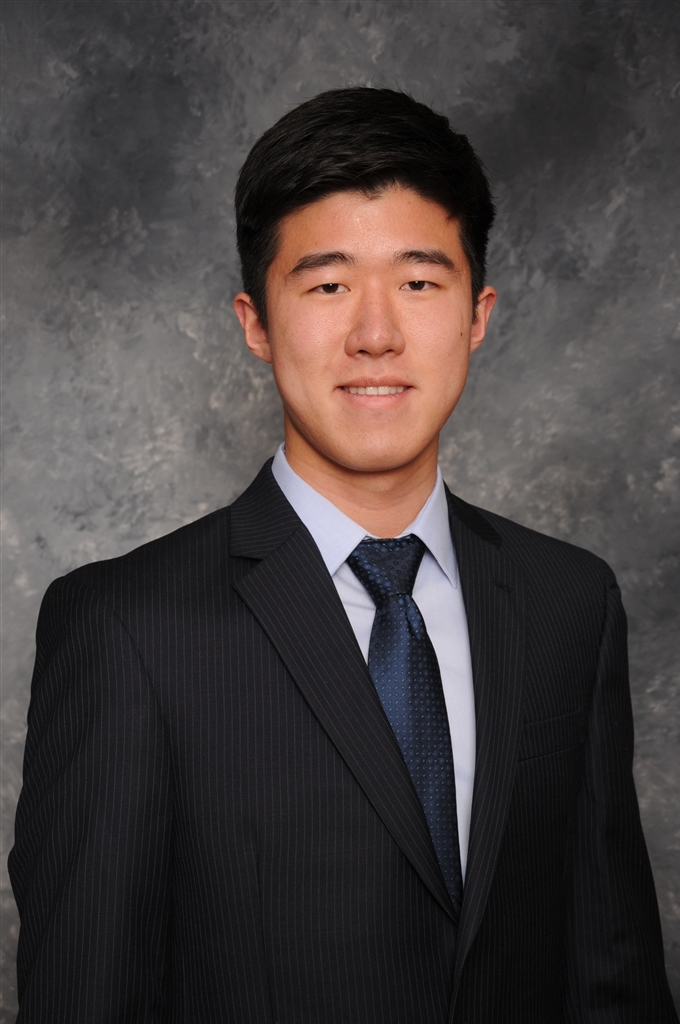}}]{Yuchen Gu}
(Graduate Student Member, IEEE) received the B.S.E.E. degree from University of Wisconsin-Madison, Madison, WI, USA, in 2019. He has been pursuing the Ph.D. degree there since 2019. He has held multiple positions on research and product development with Skyworks Solutions, Cedar Rapids, IA, Qualcomm Inc, San Diego, CA, Google Hardware, Mountain View, CA. His research interests include RF/microwave applications in medicine, and machine learning enabled OTA measurement techniques. 
\end{IEEEbiography}

\begin{IEEEbiography}[{\includegraphics[width=1in,height=1.25in,clip,keepaspectratio]{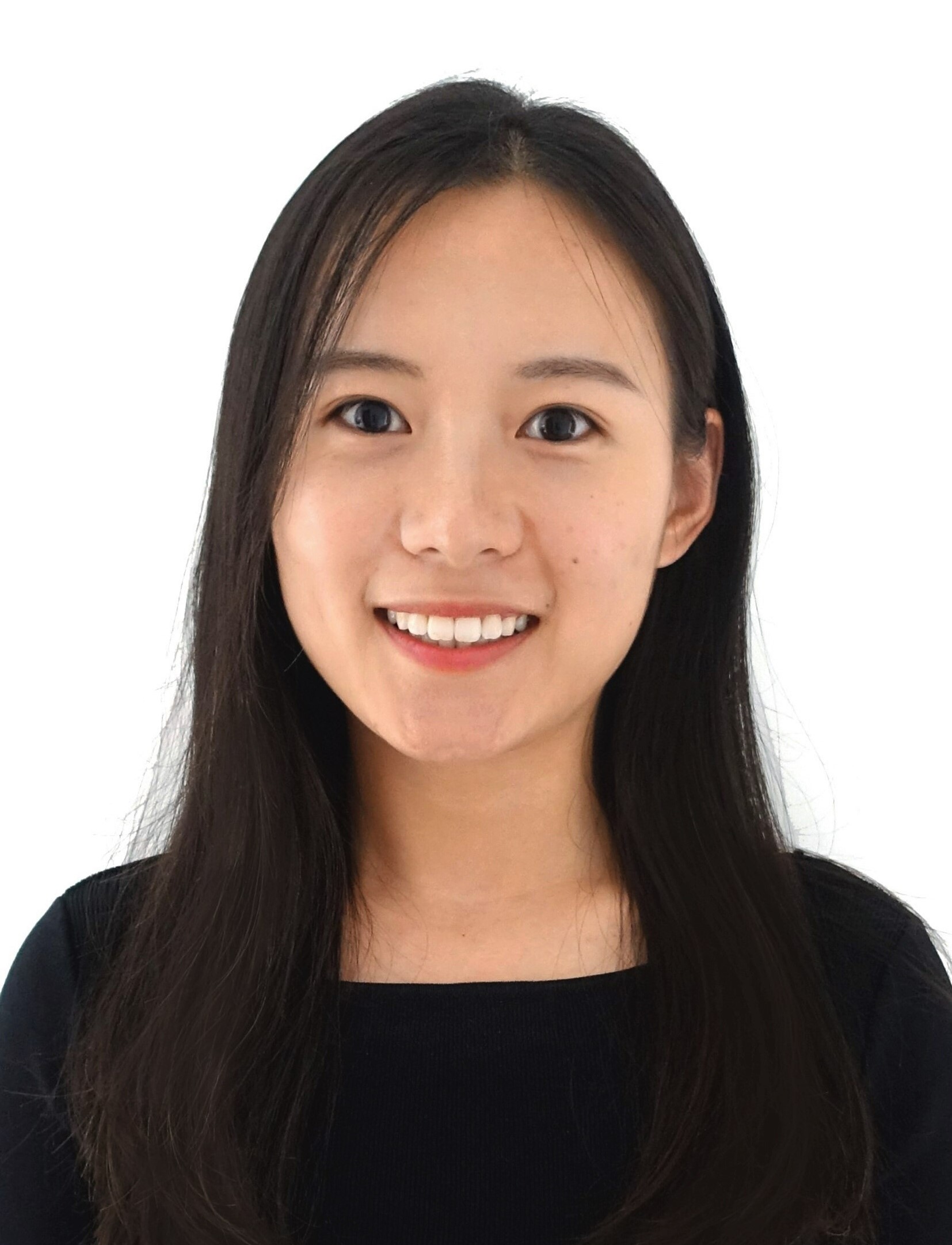}}]{Hai-Han Sun} (Member, IEEE)
received her bachelor’s degree in electronic information engineering from Beijing University of Posts and Telecommunications, Beijing, China, in 2015, and the Ph. D. degree in engineering from the University of Technology Sydney, Australia, in 2019. From 2019 to 2023, she was a Research Fellow at Nanyang Technological University, Singapore. She is currently an assistant professor in the Department of Electrical and Computer Engineering at the University of Wisconsin-Madison. Her research interests include ground-penetrating radar, base station antenna, electromagnetic sensing, and non-destructive testing. 
\end{IEEEbiography}

\begin{IEEEbiography}[{\includegraphics[width=1in,height=1.25in,clip,keepaspectratio]{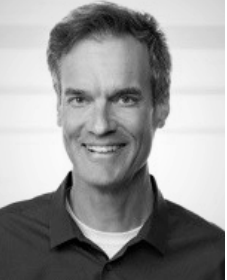}}]{Daniel W. van der Weide } (Fellow, IEEE) received the B.S.E.E. degree from The University of Iowa, Iowa City, IA, USA, in 1987, and the master’s and Ph.D. degrees in electrical engineering from Stanford University, Stanford, CA, USA, in 1989 and 1993, respectively. He held positions at the Lawrence Livermore National Laboratory, Livermore, CA, USA, Hewlett-Packard, Santa Rosa, CA, USA, Motorola, Schaumburg, IL, USA, and Watkins-Johnson Company, Palo Alto, CA, USA. From 1993 to 1995, he was a Post-Doctoral Researcher with the Max-Planck-Institute for Solid State Research, Stuttgart, Germany. He then joined the Department of Electrical and Computer Engineering, University of Delaware, Newark, DE, USA, as an Assistant Professor, where he was promoted to Associate Professor while founding and directing the Center for Nanomachined Surfaces. In 1999, he moved to the University of Wisconsin–Madison, Madison, WI, USA, where he is currently a Full Professor of electrical and computer engineering. He is also a Co-Founder of Neuwave Medical Inc., Madison (acquired by Johnson \& Johnson in 2016), Optametra LLC, Verona, WI, USA (acquired by Tektronix in July 2011), Tera-X LLC, Verona, and, most recently, Elucent Medical, Eden Prairie, MN, USA, and Accure, Madison. 
Dr. Weide received the PECASE Award from the National Science Foundation in 1997, the DARPA Ultra Electronics Program Outstanding Individual Technical Achievement Award in 1997, and the Young Investigator Program Award from the Office of Naval Research in 1998. 
\end{IEEEbiography}

\end{document}